\documentclass[11pt]{article}
\usepackage{amsfonts,amssymb,amsthm,amsmath,graphicx,bm}
\usepackage[usenames]{color}
\usepackage{psfrag}

\definecolor{Red}{cmyk}{0,1,1,0}

\newcommand{\R}{{\mathord{\mathbb R}}}

\newcommand{\N}{{\mathord{\mathbb N}}}

\newcommand{\E}{{\mathord{\mathbb E}}}

\newcommand{\HH}{\mathcal{H}}

\newcommand{\FF}{\mathcal{F}}

\newcommand{\hh}{\mathfrak{h}}

\newtheorem{lemma}{Lemma}
\newtheorem{theorem}[lemma]{Theorem}
\newtheorem{remark}[lemma]{Remark}
\newtheorem{proposition}[lemma]{Proposition}

\numberwithin{equation}{section}
\numberwithin{lemma}{section}

\def\bbone{{\mathchoice {\rm 1\mskip-4mu l} {\rm 1\mskip-4mu l}
{\rm 1\mskip-4.5mu l} {\rm 1\mskip-5mu l}}}

\begin{document}

\title{Analyticity of The Ground State Energy For Massless Nelson Models}

\author{Abdelmalek Abdesselam${ }^{1}$, David Hasler${ }^{2,}$\footnote{On leave from:
Department of Mathematics, College of William and Mary, Williamsburg VA, 23187-8795, USA.}}
\maketitle

\begin{center}
{\it ${ }^1$ Department of Mathematics,
P. O. Box 400137,
University of Virginia,
Charlottesville, VA 22904-4137, USA }\\
email: \texttt{aa4cr@virginia.edu} \\
{\it ${ }^2$ Department of Mathematics,
Ludwig Maximilans University,
Theresienstrasse 39,
Munich, D-80333, Germany }\\
email: \texttt{hasler@math.lmu.de}
\end{center}

\begin{abstract}
We show that the ground state energy of the translationally invariant Nelson model,
describing a particle coupled to a relativistic field of massless bosons,
is an analytic function of the coupling constant and the total momentum.
We derive an explicit expression for the ground state energy which is
used to determine  the effective  mass.
\end{abstract}

\section{Introduction and  Results}

We consider the so-called
translationally invariant massless Nelson model in $d$ dimensions,
which  describes a
quantum mechanical particle interacting with a quantized field of
relativistic massless bosons  \cite{nel64}. Both, particle and field move
in $d$ dimensions and we assume that $d \geq 3$.

The Hamiltonian of this model will be denoted by $H_\lambda$, with  $\lambda \in \R$
denoting the coupling constant. The generators of translations in $\R^d$, which are also known as operator of total momentum,
commute with the Hamiltonian.
This yields a
direct integral decomposition of the Hamiltonian
\begin{equation} \label{eqcor:directintegral}
H_\lambda \cong \int^{\oplus}_{\R^d}  H_\lambda(P)  d^d P  ,
\end{equation}
with  fiber Hamiltonian $H_\lambda(P)$, \cite{reesim78}.
This paper is devoted to the  properties of the ground state  energy
\begin{equation} \label{eqcor:defofgs}
E_\lambda(P) := \inf \sigma( H_\lambda(P) )
\end{equation}
as a function of the coupling constant $\lambda$ and the total momentum $P$.
Strictly speaking, one should refer to   \eqref{eqcor:defofgs} as the  minimal energy,
since in three dimensions the fiber Hamiltonian
does not have a ground state  \cite{fro73} (for non-relativistic QED see also~\cite{hasher08}).
On the basis of heuristic physical principles, $E_\lambda(\cdot)$ should
be  twice continuously differentiable in a neighborhood of the origin, with positive
second  derivative.  Establishing these properties  is technically
difficult due to the fact that the ground state energy  is at the bottom of
the continuous spectrum.
By rotation invariance $E_\lambda(P)$ only depends
on the absolute value of $P$.
Using functional integrals one can show    that $E_\lambda(\cdot)$ has a global minimum  at the origin,   \cite{gro72}.
Fr\"ohlich showed in his thesis   that the gradient of  $E_\lambda(\cdot)$   is Lipschitz continuous \cite{fro73,fro74}, see also \cite{pizzo05}.
Using renormalization \cite{bacchefro07,che08} or iterated perturbation theory \cite{fropiz10}
it was shown  for non-relativistic QED,
that $E_\lambda(\cdot)$ is in a neighborhood of  the origin a  $C^2$--function
with positive second derivative;  the same methods and results should  apply for the Nelson model as well.
However, these results  on the  regularity do not go beyond $C^2$.

In this paper it is shown  that $E_\lambda(P)$  is   in a neighborhood of zero a real analytic  function
of $\lambda$ as well as  $P$.
We  provide  explicit analytic  expansions of the ground state energy
in powers of  $\lambda$.
From this expansion it follows that on
 any closed subinterval of $(-1,1)$ and for sufficiently
small values of the coupling constant,
the ground state energy $E_\lambda(\cdot)$ is a  $C^2$--function with positive second derivative.
Furthermore, we derive a convergent small-coupling expansion for
the
effective mass
\begin{equation}\label{eqcor:defofeffmass}
m_{\rm eff}(\lambda) := \frac{1}{\partial_{P_\nu}^2  E_\lambda(P) |_{P=0}}    , \quad \nu=1,...,d ,
\end{equation}
in powers of $\lambda$ (see \cite{spohnmass}
for  the study of the effective mass for the polaron
and \cite{haisei02,hirspo05,bacchefro07} for non-relativistic QED).
As is well known, Nelson's model studied in this article is very
close to the polaron model which is of considerable physical interest
(see \cite{DevreeseA} for a recent review). The literature on the polaron is enormous
and a considerable part of it is concerned with the calculation of this effective
mass (see \cite{GerlachL,AlexandrouR} for reviews on this particular problem).
Some of the references most relevant to our approach for the calculation of $m_{\rm eff}(\lambda)$
are \cite{KholodenkoF,BogoliubovP,Smondyrev,GerlachLS,Rosenfelder}.
In view of the importance of such calculations, we therefore present
an explicit combinatorial convergent expansion for the effective mass, in the
Nelson model case.

The method used in this article is a variant of the one introduced in~\cite{AspinB} for the treatment
of the massless spin-boson model. Our method
exploits the well known formula which relates the minimal energy to the vacuum expectation
of the semigroup generated by the Hamiltonian, i.e.,
 \begin{equation} \label{eq:eqcor:heatker}
E_\lambda(P)   = \lim_{T \to \infty} - \frac{1}{T} \log  ( \Omega , e^{- TH_\lambda(P)} \Omega)  .
\end{equation}

Using functional integration and a forest interpolation
formula due to Brydges, Kennedy, Abdesselam, and Rivasseau (BKAR formula)
we analyze the right hand side of \eqref{eq:eqcor:heatker}.
The functional integral representation of the quantity $( \Omega , e^{- TH_\lambda(P)} \Omega)$
essentially provides a mapping of the original model to a statistical mechanics 
model of interacting intervals on the real line.
The right-hand side of \eqref{eq:eqcor:heatker} then becomes the infinite volume limit of the pressure
for this gas of intervals. The latter is then studied using a cluster expansion obtained from the BKAR
formula which decouples the two-body interactions of these intervals in a minimal way.
Namely, the formula can be viewed as a resummation in terms of trees of a sum over graphs with edges
corresponding to the two-body interactions between intervals. 
This yields an explicit expansion of the minimal energy.

We believe that the analyticity part of our result could alternatively
be obtained using techniques based on operator theoretic renormalization,
\cite{bacfrosig98,bacchefro07,che08,grihas09}.  However, the explicit expansion which we derive
seems to be specific to the statistical mechanics methods employed in this paper.

To illustrate the methods used in this paper
we shall first consider a harmonic oscillator
coupled to a relativistic field of massless bosons. This
can be viewed as a didactic introduction to the full-fledged expansion in the translation invariant
case.
It will be shown that also for that harmonic oscillator model the ground state energy
is an analytic function of the coupling constant $\lambda$ and  an explicit formula
for the ground state energy will be derived.
We note that the  analyticity part of this result can in fact be deduced   for
a larger class of confining potentials, not just a harmonic oscillator potential,
using  the result of Griesemer and Hasler in \cite{grihas09}.  That result  is
based on operator theoretic renormalization and   requires
the Hamiltonian to satisfy  a mild IR-condition.
By  working in an IR-regular representation of the CCR, \cite{arai01,sasaki05},
the Hamiltonian has sufficiently regular IR-behavior, for  \cite{grihas09} to be applicable, while  its minimal energy remains unchanged (we thank the referee for pointing this out to us).

It would be interesting to further investigate whether  the methods and
estimates used in this  paper can be extended to  the renormalized Nelson model, i.e.,
where  the ultraviolet cutoff is removed according to the procedure outlined in \cite{nel64}.

Below we introduce the model and state the results.
The bosonic Fock space is given by
$$
\mathcal{F} = \mathbb{C} \oplus \bigoplus_{n=1}^\infty \hh^{\otimes_s^n} ,
$$
with vacuum vector $\Omega = (1,0,0,...)$ over the Hilbert space $\hh = L^2(\R^d)$.
By  $a^*(k)$ and $a(k)$, $k \in \R^d$,  we denote the usual bosonic  creation and annihilation operators
in $\FF$ satisfying canonical commutation relations
\begin{equation} \label{eq:CCR1}
 [a(k), a(k')]=[a^*(k),a^*(k')]=0 , \quad
 [a(k), a^*(k')] = \delta(k-k') ,
 \end{equation}
and
\begin{equation} \label{eq:CCR2}
a(k) \Omega = 0 .
\end{equation}
Equations \eqref{eq:CCR1} and \eqref{eq:CCR2} are understood in the sense of distributions over $\R^d$.
We shall consider massless bosons with dispersion relation $\omega(k) = |k|$. The operator of the free field
is given by the self-adjoint operator
$$
H_f := \int_{\R^d}    \omega(k) a^*(k) a(k)  d^d k       .
$$
We define the real Hilbert space $H_{-1/2}$ to be  the completion of $\mathcal{S}_{\rm real}(\R^d)$
with respect to  the inner product
$$
( f , g )_{-1/2} :=  \int_{\R^d}  \frac{ \overline{\widehat{f}(k)} \widehat{g}(k) }{2\omega(k) } {d^d k}  .
$$
For $f \in H_{-1/2}$ we define the field operator
$$
{\phi}(f) := \int_{\R^d} \frac{ d^d k   }{\sqrt{2 \omega(k)}} ( \widehat{f}(k) a(k) + \widehat{f}(-k) a^*(k) )  .
$$
It is well known, see for example \cite{bacfrosig98}, that the following estimate follows as a consequence of
the Cauchy-Schwarz inequality. For all $\varphi \in D(H_f^{1/2})$
\begin{equation} \label{eq:boundonvarphi}
\| \phi(f) \varphi \| \leq 2 \| \widehat{f} / {\omega} \|_\hh \| H_f^{1/2} \varphi \| + \| \widehat{f}/  \sqrt{\omega}   \|_\hh   \| \varphi \| ,
\end{equation}
where  $\| \cdot \|_\hh$ denotes the norm of $\hh$.

Let   $\rho \in  H_{-1/2}$ and  $\widehat{\rho}/\omega  \in \hh$.
For  $x \in \R^d$   define   $\rho_x(\cdot) := \rho(\cdot + x)$.
The Hamiltonian of the translationally invariant Nelson model  is given by
$$
H_\lambda := - \frac{1}{2} \Delta_x + H_f  + \lambda \phi(\rho_x)         ,
$$
and acts  in the Hilbert space $\HH:= L^2(\R^d)\otimes \FF$.
The Hamiltonian is invariant under translations and
thus commutes with $- i \nabla_x + P_f$ the operator of total momentum, where
$$
P_f :=  \int_{\R^d}   k a^*(k) a(k) d^d k .
$$
This yields the  direct integral decomposition \eqref{eqcor:directintegral} with fiber Hamiltonian
$$
H_\lambda(P) = \frac{1}{2}( P - P_f )^2  + \lambda \phi(\rho_0) + H_f ,
$$
acting  in $\FF$.
By \eqref{eq:boundonvarphi} one sees  that $\phi(\rho_x)$
is infinitesimally small with respect to $H_f$.
Hence  $H_\lambda$ and $H_\lambda(P)$ are  self-adjoint on the natural domain of $H_0$ and $H_0(P)$, respectively \cite{reesim2}.
We now state the main result of this paper.

\begin{theorem} \label{thm:main2} Assume that $\widehat{\rho} \geq  0$ a.e.. Then $E_\lambda(P)$
is a real analytic function of $\lambda$ and $P$ on the set
\begin{equation}\label{eq:condonlambdaP}
|\lambda| < \frac{1}{2}\times (1-|P|)^{-\frac{3}{2}}\times
\left(
\int_{\mathbb{R}^d} d^dk\ \frac{|\widehat{\rho}(k)|^2}{|k|^2}
\right)^{-\frac{1}{2}}\ .
\end{equation}
The expansion coefficients are given in Equation \eqref{eq:explicitexpansiontr}.
\end{theorem}

\begin{proof}
First assume that $\widehat{\rho} > 0$ a.e.. In that case
 the conclusion of Theorem  \ref{thm:main2} follows
as a consequence of Theorems  \ref{thm:parttrans1} and   \ref{thm:parttrans},
Proposition   \ref{eq:transmainexp}, and Inequality \eqref{eq:defoflambda01}.
To extend it to  the case
$\widehat{\rho} \geq 0$, we choose
 an approximating sequence $({\rho}_n)_{n \in \N}$ in $H_{-1/2}$, with
$\widehat{\rho}_n \geq \widehat{\rho}_{n+1} > 0$,  such that for $s=1,2$ the following limit converges in  $\hh$,
$$
 \widehat{\rho}_n \omega^{-\frac{s}{2}} \stackrel{n \to \infty}{\longrightarrow}  \widehat{\rho} \omega^{-\frac{s}{2}} . %\label{eq:convcoup}
 $$
 Let $E_{n,\lambda}(P) := \inf \sigma ( H_0(P) + \lambda \phi(\rho_n))$. Then by
\eqref{eq:boundonvarphi} %and \eqref{eq:convcoup}
it follows that  $$\lim_{n \to \infty} E_{n,\lambda}(P) = E_\lambda(P).$$ On
the other hand, in view of  \eqref{eq:defoflambda01} we can
  interchange the limit $n \to \infty$ with the infinite summation
in the expansion  \eqref{eq:explicitexpansiontr} of  $E_{n,\lambda}(P)$. Now each
coefficient in that expansion converges as $n \to \infty$, by dominated convergence.
\end{proof}

\begin{remark}
The assumption $\widehat{\rho} \geq  0$ in Theorem \ref{thm:main2} includes the case which
is of physical interest.
Nevertheless, one could relax this assumption for  example  as follows.
Let   $\chi_\sigma(k) = 1_{[\sigma,\infty)}(|k|)$. The assertion of Theorem~\ref{thm:main2} holds,
provided  for all $\lambda$ and $P$ satisfying  \eqref{eq:condonlambdaP} the
Hamiltonian  $ \frac{1}{2}( P - P_f )^2  +   \lambda\phi([\check{\chi}_\sigma \ast \rho])+ H_f $
has for any  $\sigma > 0$ a ground state which has a nonzero overlap with  the vacuum.
\end{remark}

\medskip

Let us now consider a particle in a  harmonic oscillator potential coupled to the quantized field.
The Hamiltonian of this model acts in $\HH$
and  is given by
$$
L_\lambda :=  H_{\rm osc}  + H_f + \lambda \phi(\rho_x)           ,
$$
where
$$
H_{\rm osc} := - \frac{1}{2} \Delta_x + \frac{1}{2}x^2  - \frac{d}{2} .
$$
By  \eqref{eq:boundonvarphi}, $L_\lambda$ is self-adjoint on the natural domain of $L_0$.
We  note   that the operator $L_\lambda$  is different from the explicitly solvable operator,
which has been investigated in  \cite{arai81}.
 The next theorem states that the minimal energy of  $L_\lambda$ is an
 analytic function of $\lambda$. The region for which we can prove
 analyticity will depend on the   quantity
\begin{equation} \label{eq:defofLambda}
 \Lambda := \sup_{n \ge 1} \left(  \int_{\mathbb{R}^d}
d^d k\  | \widehat{\rho}(k)|^2 |k|^{n - 2}  \right)^{1/n} .
\end{equation}
\begin{theorem} \label{thm:main1}  The infimum of the spectrum
$
E_{\lambda} := \inf \sigma(  L_\lambda)
$
is  a real  analytic function of $\lambda$ for
$$
|\lambda| <  ( 2 e \Lambda^2)^{-\frac{1}{2} } \ .
$$
The expansion coefficients are given in Equation \eqref{eq:explicitexpansionho}.
\end{theorem}

Theorem \ref{thm:main1} will follow
as a consequence of Theorem  \ref{thm:parttrans2}, Theorem \ref{thm:partho}, and
Subsection \ref{sec:expho}.

\smallskip

\section{Positivity Preserving Representations}

In this section we justify \eqref{eq:eqcor:heatker}, which will be the content of  Theorem  \ref{thm:parttrans1},
and we justify an analogous formula for the harmonic oscillator potential, which will be  the content of Theorem  \ref{thm:parttrans2}.
The following lemma will be used in  the proof of Theorem  \ref{thm:parttrans1} and Lemma \ref{lem:spec2}.

\begin{lemma}  \label{lem:spec} Let $H$ be a self-adjoint operator on a Hilbert space
$\HH$.
 For $\psi \in \HH$ we have
$$
\lim_{T \to \infty} - \frac{1}{T} \log(\psi, e^{-TH} \psi ) = \inf {\rm supp}\ \mu_\psi
$$
where  $\mu_\psi$ denotes the spectral measure of $\psi$ with respect to the operator
$H$.
\end{lemma}
\begin{proof}
Let $E_\psi = \inf {\rm supp}\ \mu_\psi$.
By the spectral theorem
$$(\psi, e^{-TH} \psi) = \int_{E_\psi}^\infty e^{- T \lambda} d \mu_\psi(\lambda) $$
 and hence for
any $\epsilon > 0$,
$$
C_\epsilon e^{- T(E_\psi + \epsilon )} \leq ( \psi , e^{- TH} \psi) \leq e^{- T E_\psi} ,
$$
where $C_\epsilon = \mu_{\psi}([E_\psi, E_\psi + \epsilon ])$. Taking the logarithm, we obtain
$$
E_\psi \leq - \frac{1}{T} \log( \psi, e^{- TH} \psi ) \leq E_\psi + \epsilon  - \frac{1}{T} \log C_\epsilon .
$$
The Lemma  now follows after taking the limit $T \to \infty$.
\end{proof}

\begin{theorem} \label{thm:parttrans1} Assume that $\widehat{\rho} >  0$ a.e. and $|P| < 1$, then \eqref{eq:eqcor:heatker} holds.
\end{theorem}
\begin{proof} W.l.o.g. assume that $\lambda >  0$ (for $\lambda =0$ notice that
$\Omega$ is the ground state of $H_0(P)$). Let  $T > 0$.
We write $\hh = \hh_{\rm real} \oplus i \hh_{\rm real}$, where $\hh_{\rm real} = L^2(\R^d ; \R )$.
We introduce $\FF(\hh_{\rm real})$ the real  Fock space over $\hh_{\rm real}$. This is a real Hilbert space.
We define a    Hilbert cone \cite{far72}  in $\FF(\hh_{\rm real})$ by
\begin{align*}
\mathcal{C} := \bigoplus_{n=0}^\infty \mathcal{C}^{(n)} , \quad
\mathcal{C}^{(n)} := \{ f \in \hh_{\rm real}^{\otimes_s^n} | (-1)^n f \geq 0 \} .
\end{align*}
It was shown in \cite{fro73,fro74} that $e^{- TH_\lambda(P)}$ is ergodic
with respect to  $\mathcal{C}$ (see also Section 3.3 in \cite{schmol04}
and Propositions 2 and 3 in \cite{far72}).   It follows  from  the proof of
Theorem XIII.44 (c)$\Rightarrow$(e) in \cite{reesim78}
 that  $e^{- TH_\lambda(P)}$ is positivity
improving.
Fix $\epsilon> 0$ and choose a nonzero function $f$ with
\begin{equation} \label{eq:fchoice1}
 0 \leq  f  \leq e^{ - \epsilon H_\lambda(P)} \Omega ,
\end{equation}
where the inequality is understood with respect to the cone $\mathcal{C}$, \cite{far72}.
Now using that $e^{- T H_\lambda(P)}$ is positivity preserving we find
$$
 - \frac{1}{T} \log( \Omega  , e^{- (T + 2 \epsilon) H_\lambda(P)} \Omega  ) \leq  - \frac{1}{T} \log( f   , e^{- T H_\lambda(P)} f   )  .
$$
Taking the limit as $T \to \infty$ we obtain using Lemma~\ref{lem:spec}
$$
\inf \sigma(H_\lambda(P)) \leq  \inf {\rm supp}\ \mu_\Omega \leq  \inf {\rm supp}\ \mu_f .
$$
 Let $\mathcal{X}$ denote the set of functions $f$ satisfying
\eqref{eq:fchoice1}.
Since $e^{- \epsilon H_\lambda(P)}$ is positivity improving, it follows that the linear span of  $\mathcal{X}$ is  dense in $\FF(\hh)$.
Hence $\inf \sigma(H_\lambda(P)) = \inf_{f \in \mathcal{X}}  \inf {\rm supp}\ \mu_f$,
 and the theorem follows using Lemma~\ref{lem:spec}.
\end{proof}

In the remaining part of this section we consider the harmonic oscillator potential and
prove the following theorem. Recall that the ground state of $H_{\rm osc}$ is  $\varphi_0(x) := \pi^{-\frac{d}{4}} e^{-\frac{x^2}{2}}$.
\begin{theorem} \label{thm:parttrans2}  The following  holds
 \begin{equation} \label{eq:parttrans2}
E_\lambda := \inf \sigma(L_\lambda) = \lim_{T \to \infty} - \frac{1}{T} \log  (  \varphi_0 \otimes \Omega    , e^{- T L_\lambda}  \varphi_0 \otimes \Omega  )  .
\end{equation}
\end{theorem}
To show  Theorem~\ref{thm:parttrans2}, we introduce the  Schr\"odinger representation of Fock space.
For the following construction of measure spaces we refer the reader to  \cite{sim74}.
By  Minlos' theorem, there
 exists a measure $d \mu$ on $Q := \mathcal{S}_{\rm real}'(\R^d)$ and a Gaussian
random process $\xi(f)$ on $Q$ indexed by
$f \in H_{-1/2}$   with mean zero and covariance
$$
\E_{d \mu}(\xi(f)\xi(g) ) = ( f , g )_{-1/2} .
$$
Moreover,
there exists  a  unitary transformation $U_\FF: \FF \to L^2(Q, d \mu)$ with
$$U_\FF \Omega = 1  , \quad   U_\FF  {\phi}(f) U^{-1}_\FF = \xi(f) .$$
Define
$$
U_p : L^2(\R^d) \to L^2( \R^d , \varphi_0^2 d^d x) , \quad   f \mapsto f / \varphi_0 ,
$$
which is   a unitary transformation of Hilbert spaces.
The map
$$
U  := U_p \otimes U_\FF: L^2(\R^d) \otimes \FF \to L^2(\R^d \times Q, \varphi_0^2 d^dx \otimes d \mu )
$$
is  unitary as well and satisfies $U ( \varphi_0 \otimes \Omega)  = 1 $.
A proof of the following lemma can be found in \cite{bacfrosig98}. For completeness we sketch the argument.
\begin{lemma} \label{thm:posho}  The operator
$
U e^{- T {L}_\lambda } U^{-1}
 $
is positivity preserving for $T \geq 0$.
\end{lemma}

\begin{proof} Define the cutoff function $\chi_N(y) := y \chi_{|y| \leq N}$.
Abbreviate $\phi := \phi(\rho_x)$ and set  $\phi_N := \phi \chi_{|\phi| \leq N}$.
Since  $\phi$ is  ${H}_f$ bounded, one can show
using  the spectral theorem that $L_0 + \lambda \phi_N$ converges in the limit
$N \to \infty$
 in strong resolvent sense to $L_\lambda$.
On the other hand
$U e^{- t  {L}_0 } U^{-1}$ is positivity preserving for $t \geq 0$ (see
 XIII.12 Example 3 in \cite{reesim78} or \S I.4. in  \cite{sim74}).
 It now follows using the continuity of the functional calculus and the Trotter product formula,
 that
\begin{eqnarray*}
\lefteqn{
 U e^{- T  {L}_\lambda  } U^{-1} }
\\
&&
= {\rm s-}\lim_{N \to \infty} \left[ {\rm s-}\lim_{m \to \infty}
\left[U e^{- T  {L}_0  / m} U^{-1} U  e^{- T \lambda   \phi_N / m} U^{-1}  \right]^m \right] .
\end{eqnarray*}
Since $U e^{- t \lambda  \phi_N  } U^{-1} = e^{- t\lambda \chi_N(\xi(\rho_x))}$ is a multiplication operator which
is positivity preserving,
 the lemma follows.
\end{proof}

Theorem~\ref{thm:parttrans2}  holds as a simple consequence of Lemma~\ref{thm:posho}  and
Lemma~\ref{lem:spec2}, below.
\begin{lemma}  \label{lem:spec2}
Let $H$ be a self-adjoint operator in the Hilbert space $L^2(M,d\mu)$, where $\mu$ is a probability measure.
If $e^{- T H}$ is positivity preserving,  then
$$
\lim_{T \to \infty} - \frac{1}{T}\log( 1 , e^{- T H} 1 ) = \inf \sigma (H) .
$$
\end{lemma}

\begin{proof} The following proof is from  \cite{lorminspo:02}. Using that  $e^{- TH}$ is positivity preserving
it follows that for $f \in L^2(M , d \mu  )$ with
\begin{equation} \label{eq:fchoice}
0 < c_1 \leq f \leq c_2 < \infty ,
\end{equation}
 the following inequalities hold
$$
c_1^2 ( 1 , e^{- TH}  1 ) \leq (  f , e^{- TH}  f )
\leq c_2^2 (  1 , e^{- TH}  1 )  .
$$
By Lemma  \ref{lem:spec} one concludes that $E_{ 1} := \inf {\rm supp}\ \mu_1 =  \inf {\rm  supp}\ \mu_f $. Let $\mathcal{X}$ denote  the set of functions of the form
\eqref{eq:fchoice}. The linear span of $\mathcal{X}$ is  dense in $L^2(M,d\mu)$.  It follows that $E_{ 1} = \inf \sigma(H)$, since otherwise $E_1 > \inf \sigma(H)$ and $\chi_{(-\infty,E_1)}(H) L^2(M,d\mu)$ would contain
a nonzero vector which is orthogonal to $\mathcal{X}$.
\end{proof}

\section{Path Integral Representation}

In this section we give path integral representations for
\begin{align}
Z_T(P) & :=  ( \Omega  ,   e^{- T H_\lambda(P)}   \Omega  ) , \label{eq:pathzt} \\
Z_T &:= ( \varphi_0 \otimes \Omega  ,   e^{- T L_\lambda }  \varphi_0 \otimes \Omega  ) . \label{eq:pathztp}
\end{align}
The path integral representation for
 \eqref{eq:pathzt} and \eqref{eq:pathztp} will be  given in Theorem~\ref{thm:parttrans} and Theorem~\ref{thm:partho}, respectively.

We use the following construction of measures, see \cite{sim74}.
We define the real Hilbert space $N$ as the completion of $\mathcal{S}_{\rm real}(\R^{d+1})$ with
respect to the inner product
$$
(f,g)_N =  2 \int_{\R} \int_{\R^d} \frac{\overline{\widehat{f}(k_0,k)}
\ \widehat{g}(k_0,k)}{k_0^2 + k^2} d^dk\ dk_0 .
$$
By the Minlos theorem,
there exists a measure $d P$  on $\mathcal{S}_{\rm real}'(\R^{d+1})$ and a Gaussian random process $\xi(f)$
on $\mathcal{S}_{\rm real}'(\R^{d+1})$, which is   indexed by $f \in N$, with mean zero and covariance
$$
\mathbb{E}_{dP} ( \xi(f) \xi(g) ) = (f,g)_N .
$$
For $f \in H_{-1/2}$ and $t \in \R$, we have $f \otimes \delta_t \in N$, and we set
$
\xi_t(f) = \xi(f \otimes \delta_t) $.
$\xi_t(f)$  is a Gaussian random process, with mean zero and covariance
$$
\mathbb{E}_{dP}( \xi_s(f) \xi_t(g) ) =
\int \overline{\widehat{f}(k)} \widehat{g}(k) \frac{1}{ 2 \omega(k)} e^{- \omega(k)|t - s|} d^d k
=
(\Omega, \phi(f) e^{- |s-t| H_f} \phi(g) \Omega)  .
$$

For  \eqref{eq:pathzt}, we introduce by
 $(b_t)_{t \geq 0} = (b_{1,t},...,b_{d,t})_{t \geq 0}$  Brownian motion starting at zero,
 i.e., a Gaussian process  with
$$\E_{db}(b_{\alpha,t} b_{\beta,s}) = \delta_{\alpha,\beta}{\rm min}(s,t). $$
The  following theorem is shown    in \cite{spohn87}.
For the convenience of the reader, we sketch a proof in the Appendix.
\begin{theorem} \label{thm:parttrans}
For  \eqref{eq:pathzt} the following equation holds
\begin{align} \label{eq:parttrans}
Z_T(P) =
 \E_{db \otimes dP }  \left[ \exp(  - \lambda \int_0^T \xi_s(\rho_{b_s})ds  + i P \cdot b_T )  \right] .
\end{align}
\end{theorem}

For \eqref{eq:pathztp}, we introduce
$q_t = (q_{1,t},...,q_{d,t})$  the oscillator
process, i.e., a Gaussian process  with
$$\E_{dp}(q_{\alpha,t} q_{\beta,s}) = \frac{1}{2}\delta_{\alpha,\beta}\exp(-|t-s|) . $$
The following theorem can be shown in the same way as the Feynman-Kac formula
was proven in \cite{simfiqp79}, see also  \cite{lorminspo:02,bethirlorminspo02} and references therein. For the convenience of
the reader we sketch a proof in the Appendix.
\begin{theorem} \label{thm:partho} For  \eqref{eq:pathztp} the following equation holds
\begin{align} \label{eq:partho}
Z_T  = \mathbb{E}_{dp \otimes dP}\left[ \exp( - \lambda \int_0^T \xi_t(\rho_{q_t})dt   )  \right]  .
\end{align}
\end{theorem}

\begin{remark}
Using  Kolmogorov's continuity theorem we can assume that $t \to  \xi_t(\rho_{b_t})$ and
$t \mapsto \xi_t(\rho_{q_t})$ are a.e. continuous, and hence the integrals in  \eqref{eq:partho}
and \eqref{eq:parttrans}    exist a.e. as  Riemann integrals.
\end{remark}

\section{Power  Series Expansion}

In this section we analyze the expansion of the r.h.s. of \eqref{eq:parttrans} and \eqref{eq:partho}.
We use the  BKAR decoupling formula to calculate the logarithm of these expansions.
Our estimates on the resulting expression will allow us to take the limits  \eqref{eq:eqcor:heatker} and \eqref{eq:parttrans2}.
In Subsection~\ref{theexpsec} we analyze  the expansion for the harmonic oscillator.
This subsection serves as a preparation for Subsection~\ref{theexpsec2}, where
the translationally invariant Nelson model will be studied.

\subsection{The Expansion for the Harmonic Oscillator}\label{theexpsec} \label{sec:expho}

Integrating out the field in  \eqref{eq:partho}, we obtain
\begin{align*}
Z_T & = \mathbb{E}_{dp}  \exp\left( \frac{1}{2}  \mathbb{E}_{dP}\left[ \lambda \int_0^T \xi_t(\rho_{q_t})dt \right]^2   \right)  \\
& = \mathbb{E}_{dp}\left[ \exp\left(   \lambda^2 \int_0^T \int_0^T W(q_s-q_t,s-t) ds dt  \right)  \right]  ,
\end{align*}
where  we introduced the notation
\begin{align} \label{eq:defofW}
W(q,t) &:=   \frac{1}{4}
 \int_{} \frac{|\widehat{\rho}(k)|^2}{|k|} e^{ i k \cdot  q }  e^{-|k||t|} d^d k .
\end{align}
Expanding the exponential in a Taylor series,  one obtains with  $g:= \lambda^2/4$,
\begin{eqnarray*}
Z_T &&=
\E_{dp}\left[ \exp\left(  {  \lambda^2 \int_0^T \int_0^T W(q_s-q_t,s-t) ds dt }
\right)  \right]  \\
&& =
1 + \sum_{n=1}\frac{\lambda^{2n}}{n!}\E_{dp}\left[
\int_0^T \int_0^T \cdots \int_0^T \int_0^T W(q_{s_1}-q_{t_1},{s_1}-t_1) \ldots\right. \\
&& \left. \quad \times
W(q_{s_n}-q_{t_n},s_n-t_n)
ds_1 dt_1 \cdots ds_n dt_n  \begin{array}{c}\ \\ \ \end{array}\right] .
\end{eqnarray*}
Note that \eqref{eq:defofW} is bounded, which   allows to interchange
integration and summation  by  dominated convergence.
Inserting  \eqref{eq:defofW}  one   finds, after integration over the oscillator process
\begin{align}
Z_T &=\sum_{n=0}^{\infty} \frac{g^n}{n!}
\int_{[0,T]^{2n}}
\prod_{j=1}^{n} ds_j
\prod_{j=1}^{n} dt_j
\int_{\mathbb{R}^{nd}}
\prod_{j=1}^{n} d^d k_j
\prod_{j=1}^{n}
\left(
e^{-|k_j||s_j-t_j|}
\frac{|\widehat{\rho}(k_j)|^2}{|k_j|}
\right)
\nonumber \\
& \times\exp\left[
-\frac{1}{2}\sum_{i,j=1}^{n} k_i\cdot k_j A_{ij}(s,t)
\right]\label{Zstart}
\end{align}
where
\[
A_{ij}(s,t)=C(s_i,s_j)-C(s_i,t_j)-C(t_i,s_j)+C(t_i,t_j)
\]
and
\[
C(u,v)=\frac{1}{2} e^{-|u-v|}
\]
is the covariance of the one-dimensional oscillator or Ornstein-Uhlenbeck process.
Clearly for any collections of times $s$ and $t$, the matrix $A(s,t)$ is symmetric and positive-semidefinite.
At this point one has achieved the mapping of the original model to a model of statistical mechanics.
Indeed, $Z_T$ can be viewed as the grand canonical partition function of a gas of intervals $[s_j,t_j]$
on the real line (of course the bounds should be reversed if $t_j<s_j$).
These intervals are constrained by a two-body interaction corresponding to the $k_i\cdot k_j A_{ij}(s,t)$ terms
in the exponential.
Our first task is to write the quantity in (\ref{Zstart}) as an exponential.
This is what cluster expansions in statistical mechanics are typically used for. What follows is
an adaptation of this general approach to the model at hand, using the BKAR formula.

We will use the notation $[n]$ for the set $\{1,2,\ldots,n\}$.
For a finite nonempty set $E$,  let us denote by $E^{(2)}$ the set of unordered pairs $\{i,j\}$, where $i$ and $j$ are
distinct elements of $E$. We will consider the space $\R^{E^{(2)}}$ of multiplets $u = (u_l)_{l\in E^{(2)}}$ indexed by
$l \in E^{(2)}$.
A graph with vertex set $E$ can be viewed as a subset of the complete graph $E^{(2)}$.
A tree on $E$ is a graph without cycles which connects $E$.
More generally, a graph without cycles is called a forest since it is a disjoint collection of trees,
some possibly trivial, i.e., reduced to a single vertex.
Let $\mathfrak{F}$ be a forest
on $E$, and let $\vec{h} = (h_l)_{l \in \mathfrak{F}}$ be a vector of real parameters indexed by the edges $l$ in the forest
$\mathfrak{F}$. To such data one associates the element $u(\mathfrak{F},\vec{h}) =
(u(\mathfrak{F},\vec{h})_l)_{l \in E^{(2)}}$ in $\R^{E^{(2)}}$ as follows.  Let $a$ and $b$ be two distinct
elements in $E$. If $a$ and $b$ belong to two distinct connected components of the forest $\mathfrak{F}$,
then $u(\mathfrak{F},\vec{h})_{\{a,b\}} = 0$. Otherwise let, by definition,
$u(\mathfrak{F},\vec{h})_{\{a,b\}} = \min_{l} h_l$ where $l$ belongs to the unique simple
path in the forest $\mathfrak{F}$ joining
$a$ to $b$.
For $E=[n]$ we define the function
\[
F(u):=\exp\left[-\frac{1}{2}\sum_{i=1}^{n} k_i^2 A_{ii}(s,t)
-\frac{1}{2}\sum_{\substack{i,j=1 \\ i\neq j}}^{n} k_i\cdot k_j\ A_{ij}(s,t)\times u_{\{i,j\}}
\right]
\]
on the set $\R^{E^{(2)}}$. Observe that
\begin{equation}
F(1) = \exp\left[
-\frac{1}{2}\sum_{i,j=1}^{n} k_i\cdot k_j\ A_{ij}(s,t)
\right] ,
\end{equation}
where by $1$ we denoted the point in $\R^{E^{(2)}}$ with all components equal to 1.
Using the BKAR decoupling/interpolation formula~\cite{BK,AR} (see~\cite{Anotes}
for a pedagogical introduction) we have
\begin{equation}
F(1) = \sum_{\substack{\mathfrak{F} \ {\rm forest}\\ {\rm on}\ E}}
\int_{[0,1]^\mathfrak{F}} d \vec{h}\ \frac{ \partial^{|\mathfrak{F}|} F}{\prod_{l \in \mathfrak{F}} \partial u_l } \left( u(\mathfrak{F},\vec{h}) \right) ,
\label{BKARout}
\end{equation}
where the sum is over all forests $\mathfrak{F}$ with vertex set $E$, the notation $d \vec{h}$ is for the Lebesgue measure on the set of parameters $[0,1]^{\mathfrak{F}}$ (here $\mathfrak{F}$ is identified with its set of edges),
the partial derivatives of $F$ are with respect to the entries indexed
by the pairs belonging to $\mathfrak{F}$, and the evaluation of these derivatives is at the $\vec{h}$ dependent interpolation points
$u(\mathfrak{F},\vec{h})$.

Note that each derivative $\frac{\partial}{\partial u_{\{i,j\}}}$, for $\{i,j\}\in\mathfrak{F}$,
produces a factor
\[
-k_i\cdot k_j\ A_{ij}(s,t).
\]
Besides, the content of the exponential now becomes
\[
-\frac{1}{2}\sum_{i=1}^{n} k_i^2 A_{ii}(s,t)
-\frac{1}{2}\sum_{\substack{i,j=1 \\ i \neq j}}^{n} k_i\cdot k_j\ A_{ij}(s,t)
u(\mathfrak{F},\vec{h})_{\{i,j\}}\ .
\]
An essential feature of the BKAR decoupling formula is that it preserves positivity.
This translates into the following lemma.

\begin{lemma}\label{Positivlem}
For all $s,t \in \R^n$, $k \in \R^{d n }$,   forest $\mathfrak{F}$ on $[n]$,  and $\vec{h} \in [0,1]^{\mathfrak{F}}$, we have
$$
\frac{1}{2}\sum_{i=1}^{n} k_i^2 A_{ii}(s,t)
+\frac{1}{2}\sum_{\substack{i,j=1 \\ i \neq j}}^{n} k_i\cdot k_j\ A_{ij}(s,t) u(\mathfrak{F},\vec{h})_{\{i,j\}}
\geq 0 \ .
$$
\end{lemma}

\begin{proof}
Since $A_{ab}(s,t)$ is positive-semidefinite, we have for any subset $O$ of $[n]$,
$$
\sum_{i,j \in O} k_i \cdot k_j\ A_{ij}(s,t)\ge 0\ .
$$
For any given $\mathfrak{F}$ and $\vec{h}$ one can find (see~\cite{Anotes}) nonnegative numbers $\lambda_1,...,\lambda_p$ satisfying
$\sum_{q=1}^p \lambda_q =1$ as well as partitions $\pi_1,...,\pi_p$ of $[n]$, such that
$$
u(\mathfrak{F},\vec{h}) = \sum_{q=1}^p \lambda_q v_{\pi_q} ,
$$
where for a partition $\pi$ of $[n]$ we have defined
$(v_\pi)_{l} := \bbone\{ \exists O\in \pi, l \subset O \}$.
Henceforth we will use $\bbone\{\cdots\}$
for the sharp characteristic function of the condition
between braces.
Now, we  obtain
\[
\frac{1}{2}\sum_{i=1}^{n} k_i^2 A_{ii}(s,t)  +
\frac{1}{2}\sum_{\substack{i,j=1 \\ i\neq j}}^{n} k_i\cdot k_j A_{ij}(s,t)
u(\mathfrak{F},\vec{h})_{\{i,j\}}\qquad\qquad\qquad
\]
\begin{align*}
\qquad & =\frac{1}{2}\sum_{i=1}^{n} k_i^2 A_{ii}(s,t)  +
\sum_{\{i,j\}\in [n]^{(2)}} k_i\cdot k_j\ A_{ij}(s,t)\ u(\mathfrak{F},\vec{h})_{\{i,j\}} \\
 & =\frac{1}{2}\sum_{i=1}^{n} k_i^2 A_{ii}(s,t)  +
\sum_{q=1}^p \lambda_q \sum_{\{i,j\}\in [n]^{(2)}} (v_{\pi_q})_{\{i,j\}}
k_i\cdot k_j\ A_{ij}(s,t)
\end{align*}
\begin{align*}
\qquad & =\frac{1}{2}\sum_{i=1}^{n} k_i^2 A_{ii}(s,t)  +
\sum_{q=1}^p \lambda_q \sum_{O\in\pi_q}
\sum_{\{i,j\}\in O^{(2)}} k_i\cdot k_j\ A_{ij}(s,t) \\
 & =\frac{1}{2}
\sum_{q=1}^p \lambda_q \sum_{O\in\pi_q}
\sum_{i,j \in O} k_i \cdot k_j A_{ij}(s,t)
\ge 0\ .
\end{align*}
\end{proof}

We now insert (\ref{BKARout}) into (\ref{Zstart})
and use Fubini's Theorem to pull the sum over the forest $\mathfrak{F}$
and the integral over the parameters $h$ out of the integrals over the $s$, $t$ and $k$'s.
This is easily justified using Lemma \ref{Positivlem}.

We therefore have
\begin{align*}
Z_T= & 1+
\sum_{n=1}^{\infty} \frac{g^n}{n!}
\sum_{\substack{\mathfrak{F}\ {\rm forest}\\ {\rm on}\ [n]}}
\int_{[0,1]^\mathfrak{F}} d\vec{h}
\int_{[0,T]^{2n}}
\prod_{j=1}^{n} ds_j
\prod_{j=1}^{n} dt_j
\int_{\mathbb{R}^{nd}}
\prod_{j=1}^{n} d^d k_j \\
 & \prod_{j=1}^{n}
\left(
e^{-|k_j||s_j-t_j|}\ \frac{|\widehat{\rho}(k_j)|^2}{|k_j|}
\right)
\times \prod_{\{i,j\}\in\mathfrak{F}}
\left(-k_i\cdot k_j\ A_{ij}(s,t)\right) \\
 & \times\exp\left[-\frac{1}{2}\sum_{i=1}^{n} k_i^2 A_{ii}(s,t)
-\frac{1}{2}\sum_{\substack{i,j=1 \\ i \neq j}}^{n} k_i\cdot k_j\ A_{ij}(s,t)
u(\mathfrak{F},\vec{h})_{\{i,j\}}
\right] \ .
\end{align*}

For given $n\ge 1$, we have a sum of the form $\sum_{\mathfrak{F}} f(\mathfrak{F})$
which can be organized according to connected components:
\begin{align*}
\sum_{\substack{\mathfrak{F} \ {\rm forest} \\ {\rm on}\ [n]}}
f(\mathfrak{F}) &= \sum_{k \geq 1} \frac{1}{k!} \sum_{J_1,..., J_k \subset [n] }
\bbone  \left\{ {\rm \ the \ } J_l \ {\rm form \ a \ partition \ of } \ [n] \ \right\} \\
& \times
\sum_{\substack{\mathfrak{T}_1,...,\mathfrak{T}_k, \\ \mathfrak{T}_l  \ {\rm tree\ on}\ J_l}}
f\left( \bigcup_{l=1}^k \mathfrak{T}_l \right)
\end{align*}
where the trees $\mathfrak{T}_l$ are summed over independently of each other.

Inserting the above identity and noting the factorization which follows from the crucial
property $u(\mathfrak{F},\vec{h})_{\{i,j\}}=0$ if $i,j$ belong to different components $J_l$, we have
\begin{align*}
Z_T = & 1 + \sum_{n=1}^\infty \frac{g^n}{n!} \sum_{k \geq 1} \frac{1}{k!}
\sum_{\substack{ p_1,...,p_k \geq 1 \\ \sum_{l=1}^k p_l = n }} \\
& \sum_{J_1,...,J_k \subset [n] } \bbone \left\{ \forall l, \, | J_l | = p_l  \right\}
\bbone \left\{    {\rm \ the \ } J_l \ {\rm form \ a \ partition \ of } \ [n]          \right\} \\
& \times \prod_{l=1}^k \Bigg\{
\sum_{\mathfrak{T}_l \ {\rm tree\ on}\ J_l}
\ \int_{[0,1]^{ \mathfrak{T}_l} } d \vec{h}_l \int_{[0,T]^{2 p_l}}
\ \prod_{j \in J_l} d s_j \prod_{j \in J_l} d t_j \\
 & \int_{(\R^{d})^{ p_l }} \prod_{j \in J_l } d^d k_j  \prod_{j \in J_l }
\left(
e^{-|k_j||s_j-t_j|}\ \frac{|\widehat{\rho}(k_j)|^2}{|k_j|}
\right)
\prod_{\{i,j\}\in\mathfrak{T}_l}
\left(-k_i\cdot k_j\ A_{ij}(s,t)\right) \\
& \times
\exp\Biggl[-\frac{1}{2}\sum_{i \in J_l}  k_i^2 A_{ii}(s,t)
-\frac{1}{2}\sum_{\substack{i,j\in J_l \\ i\neq j}}^{n} k_i\cdot k_j\ A_{ij}(s,t) u(\mathfrak{T}_l,\vec{h}_l)_{\{i,j\}}
\Biggr] \Bigg\}\ .
\end{align*}
Note that we also used the fact
$u(\mathfrak{F},\vec{h})_{\{i,j\}}=u(\mathfrak{T}_l,\vec{h}_l)_{\{i,j\}}$
for a pair $\{i,j\}$ in $J_l$. Indeed, it follows from the definitions that the computation of
this entry of the interpolation point is purely local
to the component $J_l$.

It is easy to see that the contribution of a component $J_l$
only depends on its cardinality $p_l$. Therefore, using the Multinomial Theorem and a trivial
relabeling of the summation/integration variables, on can write
\begin{align*}
Z_T = & 1 + \sum_{n=1}^\infty \frac{1}{n!} \sum_{k \geq 1} \frac{1}{k!}
\sum_{\substack{ p_1,...,p_k \geq 1 \\ \sum_{l=1}^k p_l = n }} \frac{n!}{p_1! \cdots p_k!} \\
& \prod_{l=1}^k \Biggl\{ g^{p_l} \sum_{\substack{ \mathfrak{T}  \ {\rm tree} \\ {\rm on}\ [p_l]} }
\ \int_{[0,1]^{ \mathfrak{T}} } d \vec{h} \int_{[0,T]^{ 2 p_l}}
\prod_{j =1}^{p_l} d s_j \prod_{j=1}^{p_l}  d t_j \\
& \int_{(\R^{d})^{ p_l }} \prod_{j=1}^{p_l}  d^d k_j \ \prod_{j=1}^{p_l}
\left(
e^{-|k_j||s_j-t_j|}\ \frac{|\widehat{\rho}(k_j)|^2}{|k_j|}
\right)
\prod_{\{i,j\}\in\mathfrak{T}}
\left(-k_i\cdot k_j\ A_{ij}(s,t)\right) \\
& \times
\exp\Biggl[-\frac{1}{2}\sum_{i=1}^{p_l}   k_i^2 A_{ii}(s,t)
-\frac{1}{2}\sum_{\substack{i,j=1 \\ i\neq j}}^{p_l} k_i\cdot k_j\ A_{ij}(s,t)
u(\mathfrak{T},\vec{h})_{\{i,j\}}
\Biggr] \Biggr\}\ .
\end{align*}
By cancelling the $n!$ and exchanging the order of the sums over $n$ and $k$
we immediately see that $Z_T$ becomes the exponential of
\begin{align}
\log Z_T:=
 & \sum_{n=1}^{\infty} \frac{g^n}{n!}
\sum_{\substack{\mathfrak{T}\ {\rm tree}\\ {\rm on}\ [n]}}
\ \int_{[0,1]^\mathfrak{T}} d\vec{h}
\int_{[0,T]^{2n}}
\prod_{j=1}^{n} ds_j
\prod_{j=1}^{n} dt_j
\int_{\mathbb{R}^{nd}}
\prod_{j=1}^{n} d^d k_j \nonumber \\
 & \prod_{j=1}^{n}
\left(
e^{-|k_j||s_j-t_j|}\ \frac{|\widehat{\rho}(k_j)|^2}{|k_j|}
\right)
\times \prod_{\{i,j\}\in\mathfrak{T}}
\left(-k_i\cdot k_j\ A_{ij}(s,t)\right) \nonumber \\
 & \times\exp\left[-\frac{1}{2}\sum_{i=1}^{n} k_i^2 A_{ii}(s,t)
-\frac{1}{2}\sum_{\substack{i,j=1 \\ i \neq j}}^{n} k_i\cdot k_j\ A_{ij}(s,t)
u(\mathfrak{T},\vec{h})_{\{i,j\}}
\right] \ .
\label{logZT}
\end{align}
Note that this last step relies on justifying the exchange of order of summation which follows from
the finiteness of
\begin{align*}
\Gamma_T := &
\sum_{n=1}^{\infty} \frac{|g|^n}{n!}
\sum_{\substack{\mathfrak{T}\ {\rm tree}\\ {\rm on}\ [n]}}
\int_{[0,1]^\mathfrak{T}} d\vec{h}
\int_{[0,T]^{2n}}
\prod_{j=1}^{n} ds_j
\prod_{j=1}^{n} dt_j
\int_{\mathbb{R}^{nd}}
\prod_{j=1}^{n} d^d k_j \\
 & \prod_{j=1}^{n}
\left(
e^{-|k_j||s_j-t_j|}\ \frac{|\widehat{\rho}(k_j)|^2}{|k_j|}
\right)
\times \prod_{\{i,j\}\in\mathfrak{T}}
\left(|k_i\cdot k_j\ A_{ij}(s,t)\right|) \\
 & \times\exp\left[-\frac{1}{2}\sum_{i=1}^{n} k_i^2 A_{ii}(s,t)
-\frac{1}{2}\sum_{\substack{i,j=1 \\ i \neq j}}^{n} k_i\cdot k_j\ A_{ij}(s,t)
u(\mathfrak{T},\vec{h})_{\{i,j\}}
\right] \ .
\end{align*}
The required bound $\Gamma_T<\infty$ will be shown further below.

In order to examine the limit $\lim_{T\rightarrow\infty}-\frac{\log Z_T}{T}$
it is better to use $\left[-\frac{T}{2},\frac{T}{2}\right]^{2n}$
instead of $[0,T]^{2n}$ for the domain of integration over times.
Now dividing by $T$ and taking this limit essentially amounts to
fixing a time say $s_1$ to be the origin and replace the domain of integration over times by $\mathbb{R}^{2n}$.
Indeed, the expression for the log in (\ref{logZT}) can be rewritten
\[
\log Z_T=\sum_{n=1}^{\infty} \int_{[-\frac{T}{2},\frac{T}{2}]^{2n}} ds_1\cdots dt_n\
K_n(s_1,\ldots,t_n)
\]
where the integrands are translation invariant in time (the sum over $\mathfrak{T}$ is now included in the
$K_n$'s).
So write
\[
K_n(s_1,\ldots,t_n)=K_n(0, s_2-s_1,\ldots,t_n-s_1)
\]
and change variables to these differences
\begin{align*}
\log Z_T = & \sum_{n=1}^{\infty} \int_{-\frac{T}{2}}^{\frac{T}{2}} ds_1
 \int_{\mathbb{R}^{2n-1}} ds_2\cdots dt_n\
K_n(0,s_2,\ldots,t_n) \\
 & \times \bbone\left\{
-\frac{T}{2}-s_1\le s_2\le \frac{T}{2}-s_1,\ldots
\right\} \ .
\end{align*}
Now change variables to $u=\frac{s_1}{T}$ to find
\begin{align*}
\frac{\log Z_T}{T}= & \sum_{n=1}^{\infty}
\int_{-\frac{1}{2}}^{\frac{1}{2}} du
\int_{\mathbb{R}^{2n-1}}
ds_2\cdots dt_n\
K_n(0,s_2,\ldots,t_n) \\
 & \times \bbone\left\{
-T\left(\frac{1}{2}+u\right)\le s_2\le T \left(\frac{1}{2}-u\right),\ldots
\right\}  .
\end{align*}
Note that the characteristic function goes to $1$ pointwise almost everywhere.
So, by the Lebesgue dominated convergence theorem, the key to taking the $T\rightarrow\infty$
limit is a bound on
\[
\sum_{n=1}^{\infty}
\int_{\mathbb{R}^{2n-1}}
ds_2\cdots dt_n\
|K_n(0,s_2,\ldots,t_n)|
\]
or, which is stronger, a bound on
\[
\Gamma=
\sum_{n=1}^{\infty}
\int_{\mathbb{R}^{2n-1}}
ds_2\cdots dt_n\
\bar{K}_n(0,s_2,\ldots,t_n)
\]
where $\bar{K}_n$ is defined in the same way as $K_n$
except one puts absolute values on the factors $-k_i\cdot k_j\ A_{ij}(s,t)$ and $g$ is replaced by its modulus.

Note that the previous manipulations using translation invariance and the change of variable
$u=\frac{s_1}{T}$
easily show that $\Gamma_T\le \Gamma\times T$.
Therefore, modulo the key estimate $\Gamma<\infty$ which is proved below,
we have shown that
\[
\lim_{T\rightarrow\infty}\frac{-\log Z_T}{T}=
-\sum_{n=1}^{\infty} \frac{g^n}{n!}
\sum_{\substack{\mathfrak{T}\ {\rm tree}\\ {\rm on}\ [n]}}
\int_{[0,1]^\mathfrak{T}} d\vec{h}
\int_{\mathbb{R}^{2n-1}}
\prod_{j=2}^{n} ds_j
\prod_{j=1}^{n} dt_j
\int_{\mathbb{R}^{nd}}
\prod_{j=1}^{n} d^d k_j
\]
\begin{align}
\qquad\qquad & \prod_{j=1}^{n}
\left(
e^{-|k_j||s_j-t_j|}\ \frac{|\widehat{\rho}(k_j)|^2}{|k_j|}
\right)
\times \prod_{\{i,j\}\in\mathfrak{T}}
\left(-k_i\cdot k_j\ A_{ij}(s,t)\right)  \nonumber \\
 & \times\exp\left[-\frac{1}{2}\sum_{i=1}^{n} k_i^2 A_{ii}(s,t)
-\frac{1}{2}\sum_{\substack{i,j=1 \\ i \neq j}}^{n} k_i\cdot k_j\ A_{ij}(s,t)
u(\mathfrak{T},\vec{h})_{\{i,j\}}
\right] \label{eq:explicitexpansionho}
\end{align}
where, by definition, we set $s_1:=0$.

\subsubsection{Proof of the Key Estimate $\Gamma<\infty$}\label{keyestsec}

We first
bound the big exponential by one,
using Lemma \ref{Positivlem}:
\[
\Gamma\le
\sum_{n=1}^{\infty} \left(\frac{|\lambda|^2}{4}\right)^n \frac{1}{n!} \sum_{\mathfrak{T}} \Gamma^{(n)}(\mathfrak{T}) ,
\]
where
\begin{align*}
\Gamma^{(n)}(\mathfrak{T}) := & \int_{\mathbb{R}^{2n-1}}
\prod_{j=2}^{n} ds_j
\prod_{j=1}^{n} dt_j
\int_{\mathbb{R}^{nd}}
\prod_{j=1}^{n} d^d k_j
\\
&
\times \prod_{j=1}^{n}
\left(
e^{-|k_j||s_j-t_j|}\ \frac{|\widehat{\rho}(k_j)|^2}{|k_j|}
\right)
\times \prod_{\{i,j\}\in\mathfrak{T}}
\left|-k_i\cdot k_j\ A_{ij}(s,t)\right| ,
\end{align*}
with $s_1=0$ by definition.
Note that the $h$ parameters have been integrated out.
Then one uses
\[
\left|-k_i\cdot k_j\ A_{ij}(s,t)\right|
\]
\begin{equation}
\le \frac{1}{2}|k_i|\cdot|k_j|
\left(e^{-|s_i-s_j|}+
e^{-|s_i-t_j|}+e^{-|t_i-s_j|}+e^{-|t_i-t_j|}
\right) \ .
\label{Asplit}
\end{equation}
Now one expands these factors so instead of $\mathfrak{T}$ one now has a sum over
refined trees $(\mathfrak{T},\mathfrak{s})$ which is the data made of a tree $\mathfrak{T}$
together
with the knowledge, $\mathfrak{s}$, of which exponential decay one chooses for each edge: ss, st, ts or tt,
that is $\mathfrak{s} \in \{ {\rm ss},{\rm st},{\rm ts},{\rm tt } \}^{\mathfrak{T}}$.
See the figure below for how this looks like.
\[
\parbox{10cm}{
\psfrag{0}{$0=:s_1$}
\psfrag{1}{$k_1$}
\psfrag{2}{$t_1$}
\psfrag{3}{$s_2$}
\psfrag{4}{$k_2$}
\psfrag{5}{$t_2$}
\psfrag{6}{$s_3$}
\psfrag{7}{$k_3$}
\psfrag{8}{$t_3$}
\psfrag{9}{$s_4$}
\psfrag{a}{$k_4$}
\psfrag{b}{$t_4$}
\psfrag{c}{$s_5$}
\psfrag{d}{$k_5$}
\psfrag{e}{$t_5$}
\psfrag{f}{$s_6$}
\psfrag{g}{$k_6$}
\psfrag{h}{$t_6$}
\psfrag{i}{$s_7$}
\psfrag{j}{$k_7$}
\psfrag{k}{$t_7$}
\psfrag{l}{$s_8$}
\psfrag{m}{$k_8$}
\psfrag{n}{$t_8$}
\psfrag{o}{$s_9$}
\psfrag{p}{$k_9$}
\psfrag{q}{$t_9$}
\includegraphics[width=8cm]{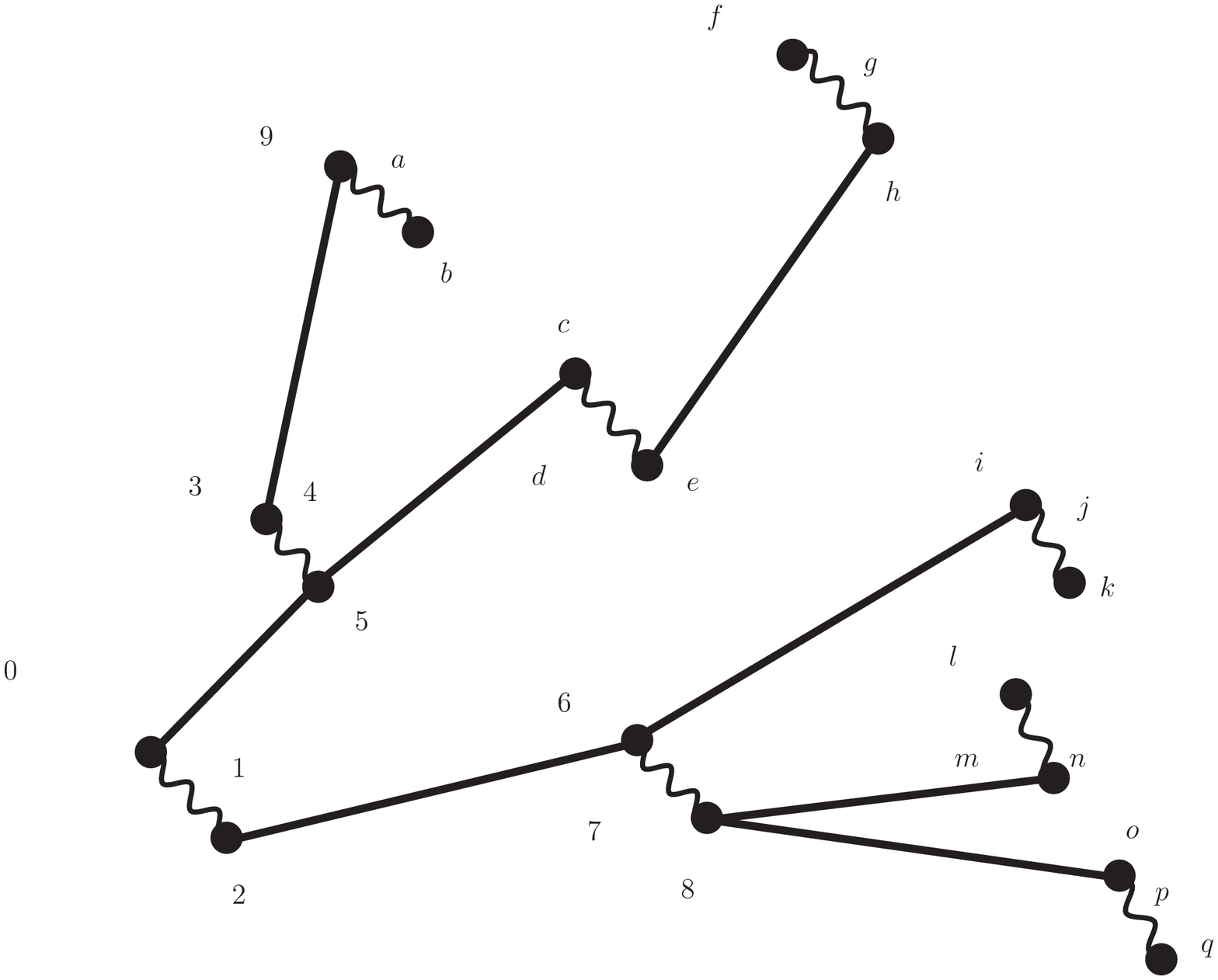}
}
\]
The hard lines correspond to the edges of the tree $\mathfrak{T}$
whereas the squiggly lines are
`internal' to its vertices.
We can now, for fixed momenta $k$, perform the integrals over the times $s$ and $t$. This integration is organized along the refined tree starting from the leaves
and progressing towards the root corresponding
to $s_1$ which is not integrated over but set equal to $0$.
For the refined tree in the
picture a possible order of integration is $t_4$, $s_4$,
$s_6$, $t_6$, $t_7$, $s_7$, $s_8$, $t_8$,
$t_9$, $s_9$, $t_3$, $s_3$, $t_5$, $s_5$,
$s_2$, $t_2$, and finally $t_1$.
The hard lines give factors of $1$ by integrating the chosen exponential
decay $\frac{1}{2} e^{-|\cdots|}$ in (\ref{Asplit}).
The squiggly lines produce $\frac{2}{|k_i|}$ factors.
Thus the refined tree $(\mathfrak{T},\mathfrak{s})$ yields a contribution to
$\Gamma^{(n)}(\mathfrak{T})$ which is
\[
\prod_{j=1}^n \left( \int_{\mathbb{R}^d}
d^d k\ 2 | \widehat{\rho}(k)|^2 |k|^{d_j - 2} \right)
\]
where $d_i$ is the degree of vertex $i$ in the tree $\mathfrak{T}$, as results from keeping track
of the $|k_i|$ factors from (\ref{Asplit}).
Then, summing over the choices of refinements $\mathfrak{s}$ for a given tree $\mathfrak{T}$,
it follows that
\[
\Gamma^{(n)}(\mathfrak{T}) \leq  4^{n-1} \prod_{j=1}^n \left(
\int_{\mathbb{R}^d}  2  | \widehat{\rho}(k)|^2 |k|^{d_j - 2} d^d k \right)
\]
where we used that a tree with $n$ vertices has $n-1$ edges.
Using the definition \eqref{eq:defofLambda} and for $n\ge 2$ we obtain
\[
\Gamma^{(n)}(\mathfrak{T}) \leq  4^{n-1}   2^n \Lambda^{2n  - 2} \ ,
\]
where we used that $d_1+ \cdots + d_n = 2 n - 2$.
The bound in the $n=1$ case only uses the hypothesis $\widehat{\rho} \omega^{-1}
\in L^2$.
As a result, we find
\[
 \sum_{\mathfrak{T}} \Gamma^{(n)}(\mathfrak{T})
 \leq 4^{n-1} 2^n  \Lambda^{ n 2 - 2} n^{n-2}
\]
by Cayley's Theorem for counting trees on $n$ vertices.
Then using $n^{n-2}\le n!\ e^{n-2}$, we finally have the desired estimate
$\Gamma<\infty$
provided $|\lambda|<\lambda_0$,
where
\begin{equation} \label{eq:defoflambda0}
\lambda_0= (2 e \Lambda^2)^{-\frac{1}{2}}\ .
\end{equation}

\subsection{Expansion for the Translation Invariant Nelson Model}   \label{theexpsec2} \label{sec:exptrans}

We integrate out the field in  \eqref{eq:parttrans} and we obtain
\begin{align*}
Z_T(P) = \E_{db}  \left(     \exp( i P b_T + \lambda^2 \int_0^T \int_0^T W(b_s-b_t,s-t)  )  \right)  .
\end{align*}
We expand the second term in the exponential in a Taylor series and  integrate out the Brownian motion.
Thereby we interchange integration and summation using dominated convergence and the boundedness of \eqref{eq:defofW}.
As a result
\begin{align}
Z_T(P) = & e^{-\frac{TP^2}{2}}\times \sum_{n=0}^{\infty} \frac{g^n}{n!}
\int_{[0,T]^{2n}}
\prod_{j=1}^{n} ds_j
\prod_{j=1}^{n} dt_j
\int_{\mathbb{R}^{nd}}
\prod_{j=1}^{n} d^d k_j \nonumber\\
& \prod_{j=1}^{n}
\left(
e^{-|k_j||s_j-t_j|-(k_j\cdot P)(s_j-t_j)}
\ \frac{|\widehat{\rho}(k_j)|^2}{|k_j|}
\right)
\nonumber \\
& \times\exp\left[
-\frac{1}{2}\sum_{i,j=1}^{n} k_i\cdot k_j\ A_{ij}(s,t)
\right]\label{Zstart2}
\end{align}
where $g=\frac{\lambda^2}{4}$,
\begin{equation}
A_{ij}(s,t)=C(s_i,s_j)-C(s_i,t_j)-C(t_i,s_j)+C(t_i,t_j)
\label{Adef1}
\end{equation}
and
\begin{equation}
C(u,v)=\min (u,v)
\label{Adef2}
\end{equation}
is the covariance of the one-dimensional Brownian motion starting at the origin.
Again, for any collections of times $s$ and $t$, the matrix $A(s,t)$ is symmetric and positive-semidefinite.

We now follow the steps in \S\ref{theexpsec}
{\it verbatim}.
The result can be summarized as follows.
Using the same notations as in \S\ref{theexpsec}, let
\begin{align*}
\Gamma:= & \sum_{n=1}^{\infty} \frac{|g|^n}{n!}
\sum_{\substack{\mathfrak{T}\ {\rm tree}\\ {\rm on}\ [n]}}
\ \int_{[0,1]^\mathfrak{T}} d\vec{h}
\int_{\mathbb{R}^{2n-1}}
\prod_{j=2}^{n} ds_j
\prod_{j=1}^{n} dt_j
\int_{\mathbb{R}^{nd}}
\prod_{j=1}^{n} d^d k_j \\
 & \prod_{j=1}^{n}
\left(
e^{-|k_j||s_j-t_j|-(k_j\cdot P)(s_j-t_j)}
\ \frac{|\widehat{\rho}(k_j)|^2}{|k_j|}
\right)
\times \prod_{\{i,j\}\in\mathfrak{T}}
\left(|k_i\cdot k_j\ A_{ij}(s,t)|\right) \\
 & \times\exp\left[-\frac{1}{2}\sum_{i=1}^{n} k_i^2 A_{ii}(s,t)
-\frac{1}{2}\sum_{\substack{i,j=1 \\ i \neq j}}^{n} k_i\cdot k_j\ A_{ij}(s,t)
u(\mathfrak{T},\vec{h})_{\{i,j\}}
\right]
\end{align*}
where, by definition, we set $s_1:=0$.
We now have the following result.

\begin{proposition} \label{eq:transmainexp}
Provided one can show the key estimate $\Gamma<\infty$,
we have that for any finite $T$, $Z_T(P)$ is the exponential of
\[
\log Z_T(P):=-\frac{TP^2}{2}+
\sum_{n=1}^{\infty} \frac{g^n}{n!}
\sum_{\substack{\mathfrak{T}\ {\rm tree}\\ {\rm on}\ [n]}}
\int_{[0,1]^\mathfrak{T}} d\vec{h}
\int_{[0,T]^{2n}}
\prod_{j=1}^{n} ds_j
\prod_{j=1}^{n} dt_j
\int_{\mathbb{R}^{nd}}
\prod_{j=1}^{n} d^d k_j
\]
\[
\prod_{j=1}^{n}
\left(
e^{-|k_j||s_j-t_j|-(k_j\cdot P)(s_j-t_j)}
\ \frac{|\widehat{\rho}(k_j)|^2}{|k_j|}
\right)
\times \prod_{\{i,j\}\in\mathfrak{T}}
\left(-k_i\cdot k_j\ A_{ij}(s,t)\right)
\]
\[
\times\exp\left[-\frac{1}{2}\sum_{i=1}^{n} k_i^2 A_{ii}(s,t)
-\frac{1}{2}\sum_{\substack{i,j=1 \\ i \neq j}}^{n} k_i\cdot k_j\ A_{ij}(s,t)
u(\mathfrak{T},\vec{h})_{\{i,j\}}
\right] \ .
\]
Furthermore, under the same hypothesis, we have
\[
\lim_{T\rightarrow\infty}\frac{-\log Z_T(P)}{T}=\frac{P^2}{2}
-\sum_{n=1}^{\infty} \frac{g^n}{n!}
\sum_{\substack{\mathfrak{T}\ {\rm tree}\\ {\rm on}\ [n]}}
\int_{[0,1]^\mathfrak{T}} d\vec{h}
\int_{\mathbb{R}^{2n-1}}
\prod_{j=2}^{n} ds_j
\prod_{j=1}^{n} dt_j
\int_{\mathbb{R}^{nd}}
\prod_{j=1}^{n} d^d k_j
\]
\[
\prod_{j=1}^{n}
\left(
e^{-|k_j||s_j-t_j|-(k_j\cdot P)(s_j-t_j)}
\ \frac{|\widehat{\rho}(k_j)|^2}{|k_j|}
\right)
\times \prod_{\{i,j\}\in\mathfrak{T}}
\left(-k_i\cdot k_j\ A_{ij}(s,t)\right)
\]
\begin{equation}
\times\exp\left[-\frac{1}{2}\sum_{i=1}^{n} k_i^2 A_{ii}(s,t)
-\frac{1}{2}\sum_{\substack{i,j=1 \\ i \neq j}}^{n} k_i\cdot k_j\ A_{ij}(s,t)
u(\mathfrak{T},\vec{h})_{\{i,j\}}
\right] \label{eq:explicitexpansiontr}
\end{equation}
where, by definition, we set $s_1:=0$. Of course, the $A$ matrix now is given by Brownian motion instead
of the oscillator process.
\end{proposition}

\subsubsection{Proof of the Key Estimate $\Gamma<\infty$}

The beginning of the argument is the same as in \S\ref{keyestsec}.
Indeed, Lemma \ref{Positivlem} holds for the new positive-semidefinite matrix $A(s,t)$.
Therefore,
\[
\Gamma\le
\sum_{n=1}^{\infty} \left(\frac{|\lambda|^2}{4}\right)^n \frac{1}{n!} \sum_{\mathfrak{T}} \Gamma^{(n)}(\mathfrak{T}) ,
\]
where
\begin{align*}
\Gamma^{(n)}(\mathfrak{T}) &:=
\int_{\mathbb{R}^{2n-1}}
\prod_{j=2}^{n} ds_j
\prod_{j=1}^{n} dt_j
\int_{\mathbb{R}^{nd}}
\prod_{j=1}^{n} d^d k_j
\\
&
\times \prod_{j=1}^{n}
\left(
e^{-|k_j||s_j-t_j|-(k_j\cdot P)(s_j-t_j)}
\ \frac{|\widehat{\rho}(k_j)|^2}{|k_j|}
\right)
\times \prod_{\{i,j\}\in\mathfrak{T}}
\left|-k_i\cdot k_j\ A_{ij}(s,t)\right| ,
\end{align*}
with $s_1=0$ and where
the $h$ parameters have been integrated out.

Now is when we part ways with the argument of \S\ref{keyestsec}.
We use, even in the case of a complex $P$, the crude bound
\begin{equation}
e^{-(k_j\cdot P)(s_j-t_j)}\le e^{|k_j||P||s_j-t_j|} \label{eq:crudebound}
\end{equation}
as well as the Cauchy-Schwarz inequality for the inner products of momenta in the factors
attached to the edges of the tree $\mathfrak{T}$. This gives the inequality
\begin{align}
\Gamma^{(n)}(\mathfrak{T}) & \le
\int_{\mathbb{R}^{2n-1}}
\prod_{j=2}^{n} ds_j
\prod_{j=1}^{n} dt_j
\int_{\mathbb{R}^{nd}}
\prod_{j=1}^{n} d^d k_j
\nonumber \\
&
\times \prod_{j=1}^{n}
\left(
e^{-|k_j|(1-|P|)|s_j-t_j|}
\ |\widehat{\rho}(k_j)|^2 |k_j|^{d_j-1}
\right)
\times \prod_{\{i,j\}\in\mathfrak{T}}
\left|A_{ij}(s,t)\right|
\label{GammaTbd}
\end{align}
where, again, $d_i$ is the degree of vertex $i$ in the tree $\mathfrak{T}$.
However, we {\it do not} break $|A_{ij}(s,t)|$ into four pieces as in (\ref{Asplit}).
Indeed the covariance $C$ now no longer has exponential decay.
The crux of our proof in the translation invariant case is that, nevertheless, the combination
$A_{i,j}(s,t)$ has a built-in decay which enforces the overlap of the time intervals
for the indices $i$ and $j$. This is a simple consequence of the independence of increments
of standard Brownian motion.
This makes the analysis similar to that for the spin-Boson model in~\cite{AspinB}.
For real numbers $\alpha$, $\beta$, let us use the notation
\begin{equation}
I(\alpha,\beta)=[\min(\alpha,\beta),\max(\alpha,\beta)]
\label{intvdef}
\end{equation}
for the closed interval with endpoints $\alpha,\beta$, regardless of their relative order.
We also use the notation $|I|$ for the length of an interval $I$.
Finally, let ${\rm sgn}(x)$ denote the sign of a real number $x$, or more precisely
\begin{equation}
{\rm sgn}(x)=\left\{
\begin{array}{cl}
1 & {\rm if}\ x>0\ ,\\
0 & {\rm if}\ x=0\ ,\\
-1 & {\rm if}\ x<0\ .
\end{array}
\right.
\label{sgndef}
\end{equation}
We can now state the following crucial fact.
\begin{lemma}
For any $s,t\in \mathbb{R}^{n}$, and $i,j\in [n]$,
\[
A_{ij}(s,t)={\rm sgn}(s_i-t_i)\times{\rm sgn}(s_j-t_j)\times
\left|I(s_i,t_i)\cap I(s_j,t_j)\right|\ .
\]
\end{lemma}
\begin{proof}
One can check the formula using (\ref{Adef1}) and (\ref{Adef2})
in all possible cases. Alternatively, one can notice that
\[
A_{ij}(s,t)=\mathbb{E}\left[
\ (B_{s_i}-B_{t_i})\ (B_{s_j}-B_{t_j})
\ \right]
\]
where $B$ is the standard Brownian motion starting at the origin.
Then one uses the independence of increments and the defining property
\[
\mathbb{E}\left[(B_{u}-B_{v})^2\right]=|u-v|\ .
\]
\end{proof}

The main tool for the estimates is the following calculation.

\begin{lemma}\label{intvintlem}
For any real number $\mu>0$, integer $p\ge 0$ and fixed real numbers $\alpha$, $\beta$,
we have
\[
\int_{\mathbb{R}^2} ds\ dt
\ |s-t|^p e^{-\mu|s-t|} \left| I(\alpha,\beta)\cap I(s,t)\right| =
\frac{2\ (p+1)!\ |\beta-\alpha|}{\mu^{p+2}}\ .
\]
\end{lemma}
\begin{proof}
It is enough to treat the case where $\alpha=0$ and $\beta>0$ which we now assume.
Denote by $I_p$ the integral on the left-hand side. Let us first consider the case $p=0$.
By trivial symmetry
\[
I_0=2 \int_{\mathbb{R}^2} ds\ dt
\ e^{-\mu|s-t|}\ \left| I(0,\beta)\cap I(s,t)\right|
\ \bbone\{s\le t\}\ .
\]
We now decompose the relevant integration domain given by the conditions
$s\le t$ and $I(0,\beta)\cap I(s,t)\neq\emptyset$ into four pieces:
\[
I_0=2(I_{0,{\rm I}}+I_{0,{\rm II}}+I_{0,{\rm III}}+I_{0,{\rm IV}})
\]
where
\begin{align*}
I_{0,{\rm I}} &= \int_{\mathbb{R}^2} ds\ dt
\ e^{-\mu|s-t|}\ \left| I(0,\beta)\cap I(s,t)\right|
\ \bbone\{s\le 0\le t\le \beta\}\ , \\
I_{0,{\rm II}} &= \int_{\mathbb{R}^2} ds\ dt
\ e^{-\mu|s-t|}\ \left| I(0,\beta)\cap I(s,t)\right|
\ \bbone\{0\le s\le t\le \beta\}\ , \\
I_{0,{\rm III}} &= \int_{\mathbb{R}^2} ds\ dt
\ e^{-\mu|s-t|}\ \left| I(0,\beta)\cap I(s,t)\right|
\ \bbone\{0\le s\le \beta\le t\}\ , \\
I_{0,{\rm IV}} &= \int_{\mathbb{R}^2} ds\ dt
\ e^{-\mu|s-t|}\ \left| I(0,\beta)\cap I(s,t)\right|
\ \bbone\{s\le 0 \ {\rm and}\ \beta\le t\}\ .
\end{align*}
Note that we ignored domain overlaps of Lebesgue measure $0$.
By `time-reversal' symmetry we have $I_{0,{\rm I}}=I_{0,{\rm III}}$.
Then by elementary calculus one obtains for these integrals the evaluations
\[
I_{0,{\rm I}}=I_{0,{\rm III}}=\frac{e^{-\mu\beta}}{\mu^3}
(e^{\mu\beta}-1-\mu\beta)\ ,
\]
\[
I_{0,{\rm II}}=\frac{e^{-\mu\beta}}{\mu^3}
(2+\mu\beta+\mu\beta e^{\mu\beta}-2 e^{\mu\beta})\ ,
\]
\[
I_{0,{\rm IV}}=\frac{\beta\ e^{-\mu\beta}}{\mu^2}\ .
\]
Then taking the total we get
\[
I_0=\frac{2\beta}{\mu^2}\ .
\]
Finally, for $p\ge 0$,
we use differentiation under the integral sign in order to show
\[
I_p=\left(-\frac{\partial}{\partial\mu}\right)^p I_0=
\left(-\frac{\partial}{\partial\mu}\right)^p \frac{2\beta}{\mu^2}
=\frac{2\ (p+1)!\ \beta}{\mu^{p+2}}\ .
\]
\end{proof}

We can now perform the integrations over times $s$ and $t$, for fixed momenta $k$, in the
right-hand side of (\ref{GammaTbd}).

\begin{lemma}
We have, in terms of the vertex degrees $d_i$ in the tree $\mathfrak{T}$,
\[
\int_{\mathbb{R}^{2n-1}}
\prod_{j=2}^{n} ds_j
\prod_{j=1}^{n} dt_j
\ \prod_{j=1}^{n}
e^{-|k_j|(1-|P|)|s_j-t_j|}
\times
\prod_{\{i,j\}\in\mathfrak{T}}
\left|A_{ij}(s,t)\right|\qquad\qquad
\]
\[
\qquad\qquad
=2^n (1-|P|)^{-3n+2}\times \prod_{j=1}^{n}\frac{d_j!}{|k_j|^{d_j+1}}\ .
\]
\end{lemma}
\begin{proof}
We use $1\in [n]$
as the root of the tree $\mathfrak{T}$ and orient
the edges towards that root.
The time variables are inductively integrated in pairs $s_j$, $t_j$
for each vertex $j$, starting from the leafs and progressing, following
the edge orientations, towards the root.
If $j\neq 1$ has vertex $i$ as a parent, then the corresponding integral
is
\[
\int_{\mathbb{R}^2} ds_j\ dt_j
\ e^{-|k_j|(1-|P|)|s_j-t_j|}\ |s_j-t_j|^{d_j-1}
\ \left|I(s_i,t_i)\cap I(s_j,t_j)\right|\qquad
\]
\[
\qquad\qquad\qquad
=\frac{2\ d_j!\ |s_i-t_i|}{[|k_j|(1-|P|)]^{d_j+1}}
\]
by Lemma \ref{intvintlem}.
The newly produced factor $|s_i-t_i|$
is dumped onto parent $i$.
Since $j$ has $d_j-1$ offsprings, the earlier integrations account for the
$|s_j-t_j|^{d_j-1}$ featuring in the vertex $j$ integral.
The root requires a special treatment since $s_1=0$ is not integrated over and
the number of offsprings is $d_1$ instead of $d_1-1$.
In this case one has to compute
\[
\int_{\mathbb{R}} dt_1\ e^{-|k_1|(1-|P|)|t_1|}\ |t_1|^{d_1}
=\frac{2\ d_1!}{[|k_1|(1-|P|)]^{d_1+1}}
\]
by a trivial Gamma Function evaluation.
\end{proof}

We now pick up the thread from Eq. (\ref{GammaTbd})
and write, using the last lemma,
\[
\Gamma^{(n)}(\mathfrak{T})  \le
2^n (1-|P|)^{-3n+2}\times \prod_{j=1}^{n} d_j!
\times
\left(
\int_{\mathbb{R}^d} d^dk\ \frac{|\widehat{\rho}(k)|^2}{|k|^2}
\right)^n\ .
\]
By Cayley's Theorem which counts trees with fixed vertex degrees
one has, for $n\ge 2$,
\[
\sum_{\mathfrak{T}} \prod_{j=1}^{n} d_j(\mathfrak{T})!
=\sum_{\substack{d_1,\ldots,d_n\ge 1 \\ \Sigma d_i=2n-2}} \frac{(n-2)!}{(d_1-1)!\cdots (d_n-1)!}
\times \prod_{j=1}^{n} d_j!
\]
where we restored the $\mathfrak{T}$ dependence in the notations.
Therefore by the arithmetic versus geometric mean inequality
\begin{align*}
\sum_{\mathfrak{T}} \prod_{j=1}^{n} d_j(\mathfrak{T})!
&= (n-2)!\times \sum_{\substack{d_1,\ldots,d_n\ge 1 \\ \Sigma d_i=2n-2}}
d_1\cdots d_n \\
 &\le (n-2)!\times \sum_{\substack{d_1,\ldots,d_n\ge 1 \\ \Sigma d_i=2n-2}}
\left[\frac{d_1+\cdots +d_n}{n}\right]^n \\
 & \le (n-2)!\times 2^n\times    \sum_{\substack{d_1,\ldots,d_n\ge 1 \\ \Sigma d_i=2n-2}} 1 \\
 & \le (n-2)! \times 2^{3n-3}\ .
\end{align*}
As a result,
\begin{align}
\Gamma \le & \frac{|\lambda|^2}{2(1-|P|)} \times
\int_{\mathbb{R}^d} d^dk\ \frac{|\widehat{\rho}(k)|^2}{|k|^2}
 \label{eq:defoflambda01}        \\
 & + \sum_{n=2}^{\infty}
 \left(\frac{|\lambda|^2}{4}\right)^n\times
 \frac{2^{4n-2}}{n(n-1)}\times (1-|P|)^{-3n+2}\times
\left(
\int_{\mathbb{R}^d} d^dk\ \frac{|\widehat{\rho}(k)|^2}{|k|^2}
\right)^n\ . \nonumber
\end{align}
Therefore, the key estimate $\Gamma<\infty$ holds as soon as
$|\lambda|<\lambda_0(P)$ with
\[
\lambda_0(P)=\frac{1}{2}\times (1-|P|)^{-\frac{3}{2}}\times
\left(
\int_{\mathbb{R}^d} d^dk\ \frac{|\widehat{\rho}(k)|^2}{|k|^2}
\right)^{-\frac{1}{2}}\ .
\]

\section{Convergent expansion for the effective mass}
We assume in this section that the cut-off function $\widehat{\rho}(k)$ is rotationally
invariant, i.e., only depends on $|k|$.
From the formula \eqref{eq:explicitexpansiontr}
for the minimal energy $E_\lambda(P)$ 
it is easy to obtain a double expansion in $\lambda$ and $P$.
All one needs to do is expand all the factors $e^{-(k_j\cdot P)(s_j-t_j)}$
as
\[
\sum_{r_j=0}^{\infty} \frac{1}{r_j!} \ 
[-(k_j\cdot P)(s_j-t_j)]^{r_j}\ .
\]
The previous estimates, and in particular the crude bound \eqref{eq:crudebound},
show that this double expansion is convergent in the domain $|\lambda|<\lambda_0(P)$.
The degree zero term in $P$ is of course $E_\lambda(0)$.
It is also easy to see that the linear term vanishes. Indeed, such
a term contains integrals of the form
\[
\int_{\mathbb{R}^{nd}}
\prod_{j=1}^{n} d^d k_j\ 
[-(k_{j_1}\cdot P) (s_{j_1}-t_{j_1})]
\times \mathcal{W}(k) 
\]
where
\[
W(k)=
\prod_{j=1}^{n}
\left(
e^{-|k_j||s_j-t_j|}
\ \frac{|\widehat{\rho}(k_j)|^2}{|k_j|}
\right)
\times \prod_{\{i,j\}\in\mathfrak{T}}
\left(-k_i\cdot k_j\ A_{ij}(s,t)\right)
\]
\[
\times\exp\left[-\frac{1}{2}\sum_{i=1}^{n} k_i^2 A_{ii}(s,t)
-\frac{1}{2}\sum_{\substack{i,j=1 \\ i \neq j}}^{n} k_i\cdot k_j\ A_{ij}(s,t)
u(\mathfrak{T},\vec{h})_{\{i,j\}}
\right]
\]
and where we suppressed the dependence on $n, \mathfrak{T}, s, t$ for lighter notation.
By rotation invariance of $\widehat{\rho}$,
one has that $W(k)$ is invariant by simultaneous rotation of all momenta $k_j$.
This implies the vanishing of the previous integral.
As for the quadratic term in $P$, it is clearly equal to
\[
\frac{P^2}{2}-\sum_{n=1}^{\infty} \frac{g^n}{n!}
\frac{1}{2}
\sum_{j_1,j_2=1}^n \sum_{\nu_1,\nu_2=1}^{d}
\sum_{\substack{\mathfrak{T}\ {\rm tree}\\ {\rm on}\ [n]}}
\int_{[0,1]^\mathfrak{T}} d\vec{h}
\int_{\mathbb{R}^{2n-1}}
\prod_{j=2}^{n} ds_j
\prod_{j=1}^{n} dt_j
\]
\[
\int_{\mathbb{R}^{nd}}
\prod_{j=1}^{n} d^d k_j\ 
k_{j_1,\nu_1} k_{j_2,\nu_2}\ P_{\nu_1} P_{\nu_2}\ (s_{j_1}-t_{j_1})(s_{j_2}-t_{j_2})
\times W(k)\ .
\]
Using the invariance property, and a change of momentum variables which is a reflection in the $\nu_1$-th
coordinate, it follows that only terms with $\nu_2=\nu_1=\nu$ survive. Besides, by rotation invariance,
these are the same for all $\nu$.
Therefore, the quadratic part is of the form $\frac{P^2}{2 m_{\rm eff}(\lambda)}$
with
\[
\frac{1}{m_{\rm eff}(\lambda)}=
1-\frac{1}{d}\sum_{n=1}^{\infty} \frac{\lambda^{2n}}{4^n\ n!}
\sum_{j_1,j_2=1}^n
\sum_{\substack{\mathfrak{T}\ {\rm tree}\\ {\rm on}\ [n]}}
\int_{[0,1]^\mathfrak{T}} d\vec{h}
\int_{\mathbb{R}^{2n-1}}
\prod_{j=2}^{n} ds_j
\prod_{j=1}^{n} dt_j
\]
\[
\int_{\mathbb{R}^{nd}}
\prod_{j=1}^{n} d^d k_j
\ (k_{j_1}\cdot k_{j_2})\ (s_{j_1}-t_{j_1})(s_{j_2}-t_{j_2})
\]
\[
\times
\prod_{j=1}^{n}
\left(
e^{-|k_j||s_j-t_j|}
\ \frac{|\widehat{\rho}(k_j)|^2}{|k_j|}
\right)
\times \prod_{\{i,j\}\in\mathfrak{T}}
\left(-k_i\cdot k_j\ A_{ij}(s,t)\right)
\]
\[
\times\exp\left[-\frac{1}{2}\sum_{i=1}^{n} k_i^2 A_{ii}(s,t)
-\frac{1}{2}\sum_{\substack{i,j=1 \\ i \neq j}}^{n} k_i\cdot k_j\ A_{ij}(s,t)
u(\mathfrak{T},\vec{h})_{\{i,j\}}
\right]\ .
\]
One can then trivially substitute into the geometric series $\frac{1}{1-x}=1+x+x^2+\cdots$
in order to obtain a convergent expansion for the effective mass $m_{\rm eff}(\lambda)$ itself.

It is important to remark that the first nontrivial term with $n=1$ is nonvoid,
since there exists a tree $\mathfrak{T}$ on the singleton $\{1\}$: the empty tree!
Indeed, if one writes the expansion for the inverse mass as
\[
\frac{1}{m_{\rm eff}(\lambda)}=1+c_2 \lambda^2+c_4\lambda^4+\cdots
\]
then one can easily extract the coefficient $c_2$
as follows.
In this case one has $n=1$, $j_1=j_2=1$, and $\mathfrak{T}=\emptyset$.
There is no integral over $h$ parameters and our formula reduces to
\[
c_2=-\frac{1}{4d}\int_{\mathbb{R}} dt_1
\int_{\mathbb{R}^{d}}d^d k_1\ 
|k_1|^2\ t_1^2\ e^{-|k_1||t_1|}\ \frac{|\widehat{\rho}(k_1)|^2}{|k_1|}
\ \exp\left[-\frac{1}{2}|k_1|^2|t_1|\right]
\]
which can readily be simplified to
\[
c_2=-\frac{1}{d}\int_{\mathbb{R}^{d}}d^d k\ 
\frac{|\widehat{\rho}(k)|^2}{|k|^2\left(1+\frac{|k|}{2}\right)^{3}}\ .
\]
Therefore the expansion for the mass is $m_{\rm eff}(\lambda)=1-c_2\lambda^2+O(\lambda^4)$
with positive $\lambda^2$ coefficient. Indeed, the physical picture is that the particle gets heavier when it is dressed by the Boson cloud.

\section*{Acknowledgements}

We are indebted to Ira Herbst for  helpful discussions.
Moreover, we are grateful for the  constructive
comments of the referee, which were incorporated in the paper.
D.H. wants  to thank  J\"urg
Fr\"ohlich for helpful conversations.
A.A. was supported in part by the National Science
Foundation under grant DMS \# 0907198.
D.H. wants to thank the University of Virginia for hospitality.

\section*{Appendix}

In this appendix we sketch proofs of Theorems \ref{thm:partho} and \ref{thm:parttrans}.
Without loss of generality we can assume that $\lambda \geq 0$.
First we address  Theorem \ref{thm:partho}.
 Define the cutoff function
$\chi_N(y) = \max (y,-N)$. Set
$$
\phi_{x,N} = \chi_N(\phi(\rho_x)) .
$$
Since $\phi(\rho_x)$ is $H_f$ bounded it follows by the spectral theorem
that $L_{\lambda,N} := H_{\rm osc} + H_f  + \lambda \phi_{x,N}$
converges in strong resolvent sense to $L_\lambda$.
Hence
\begin{align*}
( \varphi_0 \otimes \Omega  , e^{- T L_\lambda}  \varphi_0 \otimes \Omega  ) &=
\lim_{N \to \infty } ( \varphi_0 \otimes \Omega  , e^{- T L_{\lambda,N}}  \varphi_0 \otimes \Omega  ) .
\end{align*}
Using the Trotter product formula, we obtain
\begin{align*}
( \varphi_0 \otimes \Omega  , e^{- T L_{\lambda,N} }  \varphi_0 \otimes \Omega  ) & =
\lim_{m \to \infty}
 ( \varphi_0 \otimes \Omega  , \left\{  e^{- T L_{0} /m}
e^{- T \lambda \phi_{x,N}/m}  \right\}^m  \varphi_0 \otimes \Omega  ) \\
&  = \lim_{m \to \infty}
 \mathbb{E}_{dp \otimes dP }\left[
 \exp( - \sum_{i=1}^m  \lambda \chi_N( \xi_{t_i}(\rho_{q_{t_i}})) \Delta t )  \right] \\
& =  \mathbb{E}_{dp \otimes dP }\left[  \exp(  -   \lambda \int_0^T
 \chi_N( \xi_{t}(\rho_{q_{t}})) d  t )  \right] ,
\end{align*}
where in the second equality we  used the FKN  formula, see \cite{sim74}, with the notation
 $\Delta t = T/m$ and $t_i = i \Delta t $. In the third equality we used dominated convergence and that
the sum converges to the integral a.e., since $ t \mapsto \xi_{t}(\rho_{q_{t}})$ is a.e.
continuous.
Now we use monotone convergence, to conclude that the right hand side converges to
$$
 \mathbb{E}_{dp \otimes dP }(  \exp(  -  \lambda \int_0^T \xi_{t}(\rho_{q_{t}}) d  t   ) ) ,
$$
as $N \to \infty$. This shows Theorem \ref{thm:partho}.

Now we sketch a proof of  Theorem \ref{thm:parttrans}. In contrast to \cite{spohn87},
the proof below uses a Gaussian measure to linearize a square and also the so-called pull-through formula.
Using an analogous argument as above we have
\begin{align*}
(\Omega  , e^{- T H_\lambda(P)} \Omega  ) &=
\lim_{N \to \infty } (\Omega , e^{- T (H_{0}(P) + \lambda \phi_{0,N} )} \Omega ) .
\end{align*}
Using the Trotter product formula, we obtain
\begin{eqnarray*}
\lefteqn{
(\Omega  ,   e^{- T (H_{0}(P) + \lambda \phi_{0,N} )}          \Omega  ) } \\
&& =
\lim_{m \to \infty}
 (\Omega  , \left\{  e^{- T H_{0}(P)/m}
e^{- T \lambda \phi_{0,N}/m}  \right\}^m \Omega ) \\
&&  = \lim_{m \to \infty} \int d \mu_m(\alpha)
(\Omega  , \prod_{j=1}^m \left\{  e^{ i \alpha_j \cdot ( P- P_f) } e^{ - T H_f /m}
e^{- T \lambda \phi_{0,N}/m}  \right\} \Omega ) \\
&&  = \lim_{m \to \infty} \int d \mu_m(\alpha)
(\Omega  ,  e^{ i \beta_m \cdot  P  } \prod_{j=1}^m \left\{  e^{ - T H_f /m}
e^{- T \lambda \phi_{\beta_j,N}/m}  \right\} \Omega ) \\
&& =
\lim_{m \to \infty}
\mathbb{E}_{db \otimes dP}\left[  e^{ i b_T \cdot P} \exp({-  \lambda \sum_{j=1}^m \chi_N( \xi_{t_j}(\rho_{b_{t_j}}))  \Delta t } )  \right]
\\
&& =  \mathbb{E}_{db \otimes dP }\left[ e^{ i b_T  \cdot P}  \exp(  -   \lambda \int_0^T \chi_N( \xi_{t}(\rho_{b_{t}})) d  t )  \right] ,
\end{eqnarray*}
where in the second equality we introduced the Gaussian measure  $d \mu_m$  on
$\R^{3m} \ni \alpha = (\alpha_1,...,\alpha_m)$ with mean zero and covariance $T/m$.
In the third equality we used the pull-through formula and the  notation
 $\beta_n = \sum_{j=1}^n \alpha_j$. In the fourth equality we used again the
 FKN  formula. The last equality follows from the a.e. continuity of paths and dominated convergence.
Now  the right hand side converges to
$$
 \mathbb{E}_{db \otimes dP }( e^{ i b_T  \cdot P}  \exp(  -   \lambda \int_0^T  \xi_{t}(\rho_{b_{t}}) d  t )  )
$$
as $N \to \infty$. For $P=0$ this follows from monotone convergence, then it follows for $P \neq 0$ by dominated convergence.
This shows Theorem \ref{thm:parttrans}.

\end{document}